\documentclass[twoside,slac_one]{revtex4}
\usepackage{graphicx}
\usepackage{fancyhdr}
\usepackage{amsmath} 
\usepackage{bm}
\usepackage{amsxtra}
\usepackage{amssymb}
\usepackage{amsthm}
\usepackage{latexsym}
\usepackage{lscape}

\begin{document}

\title{Recent BaBar Studies of Bottomonium States}

%

\author{V. Ziegler}
\affiliation{SLAC National Accelerator Laboratory, Stanford, CA, USA}

\begin{abstract}
We present studies in bottomonium spectroscopy carried out with samples of $(98\pm1)$ million $\Upsilon(2S)$
and $(121\pm1)$ million $\Upsilon(3S)$ events recorded with the BaBar detector at the PEP-II asymmetric-energy 
$e^+ e^-$ collider at SLAC.  
Precise measurements of the branching fractions for $\chi_{b1,2}(1P, 2P)\rightarrow \gamma \Upsilon(1S)$ 
and $\chi_{b1,2}(2P)\rightarrow \gamma \Upsilon(2S)$ transitions, and searches for radiative decay to the 
$\eta_b(1S)$ and $\eta_b(2S)$ states are performed using photons that have converted into an $e^+ e^-$ pair.
In addition, we summarize the results of a search for the spin-singlet partner of the $\chi_{bJ}(1P)$ triplet, the $h_b(1P)$ state 
of bottomonium, in the transitions $\Upsilon(3S)\rightarrow \pi^0 h_b$ and $\Upsilon(3S)\rightarrow \pi^+\pi^- h_b$.

\end{abstract}

\maketitle

\thispagestyle{fancy}


\section{Introduction}

Bottomonia are the heaviest of the $q\bar{q}$ bound states.  
Decays within the bottomonium family of states occur via $\pi^0$, $\eta$, $\omega$ or di-pion emission, or by electric dipole 
transitions or magnetic dipole transitions.  
Electromagnetic transitions between the energy levels of the bottomonium spectrum can be calculated in the quark model and are an important tool in 
understanding the internal structure of $b\bar{b}$ bound states.
In particular, the measurement of the hyperfine mass splittings between triplet and singlet states is paramount to understanding the role 
of spin-spin interactions in quarkonium models and in testing QCD calculations.
In the non-relativistic approximation, the hyperfine splitting is proportional to the square of the wave function at the origin,
which is expected to be non-zero only for $L=0$,
where $L$ is the orbital angular momentum quantum number of the $q\bar{q}$ system.  

In particular, the mass splitting between the $\Upsilon(1S)$ and the $\eta_b(1S)$ is a key
ingredient in many theoretical calculations. The value measured by BaBar, $m(\Upsilon(1S))-m(\eta_b(1S))=69.3\pm2.8$ MeV/$c^2$~\cite{babar_etab1,babar_etab2} and subsequently by 
CLEO~\cite{cleo_etab} is larger than most predictions based on potential models~\cite{QWG-YR}, but in reasonable agreement with predictions
from lattice calculations~\cite{Gray:2005ur}.  
According to ref.~\cite{ref:A1}, the shift in measured $\eta_b(1S)$ compared to QCD predictions might be explained by the mixing of 
the $\eta_b$ with a CP-odd Higgs scalar.  To test this model, a measurement of the $\eta_b$ width is essential, and this was one of the BaBar 
motivations for searching for the $\eta_b$ using photon conversions as described in Sec. 2.

For $L=1$, the splitting between the spin-singlet (${}^1P_1$) and the spin-averaged triplet state ($\langle {}^3P_J\rangle$)
is expected to be $\Delta M_{\rm HF}=M({}^3P_J)-M({}^1P_1)\sim 0$.
The ${}^1 P_1$ state of  bottomonium, the $h_b(1P)$,
is the axial vector partner of the $P$-wave $\chi_{bJ}(1P)$ states.
Its expected mass, computed as the spin-weighted center of gravity of the $\chi_{bJ}(1P)$ states~\cite{ref:PDG}, is 9899.87$\pm$ 0.27 MeV/$c^2$.
Higher-order corrections might cause a small deviation from this value, but a
hyperfine splitting larger than 1 MeV/$c^2$ might be
indicative of a vector component in the confinement potential~\cite{ref:Rosner2002}.
The hyperfine splitting for the charmonium ${}^1 P_1$ state $h_c$ is measured by the BES and CLEO experiments~\cite{ref:BEShc,ref:CLEOhc0,ref:CLEOhc}
to be $\sim$0.1 MeV/$c^2$.
An even smaller splitting is expected for the much heavier bottomonium system~\cite{ref:Rosner2002}.

The BaBar~\cite{ref:babar} experiment collected data at the narrow $\Upsilon(nS)$ resonances in order to carry out detailed 
studies in bottomonium spectroscopy\footnote{These samples correspond to 14 fb$^{-1}$ and 30 fb$^{-1}$ of $\Upsilon(2S)$ and 
 $\Upsilon(3S)$ data, respectively. }.

We summarize the results of a study of radiative bottomonium transitions using $\gamma\rightarrow e^+e^-$ conversions, and of 
searches for the $h_b(1P)$ state in  $\Upsilon(3S)\rightarrow \pi^+\pi^- h_b(1P)$ decay~\cite{babar_dipihb} 
and  $\Upsilon(3S)\rightarrow \pi^0 h_b(1P)$ decay~\cite{babar_pi0hb}.  

\section{Study of radiative bottomonium transitions using  \boldmath$\gamma\rightarrow e^+e^-$ conversions}

\begin{figure}  
\begin{center}
\includegraphics[scale=0.38]{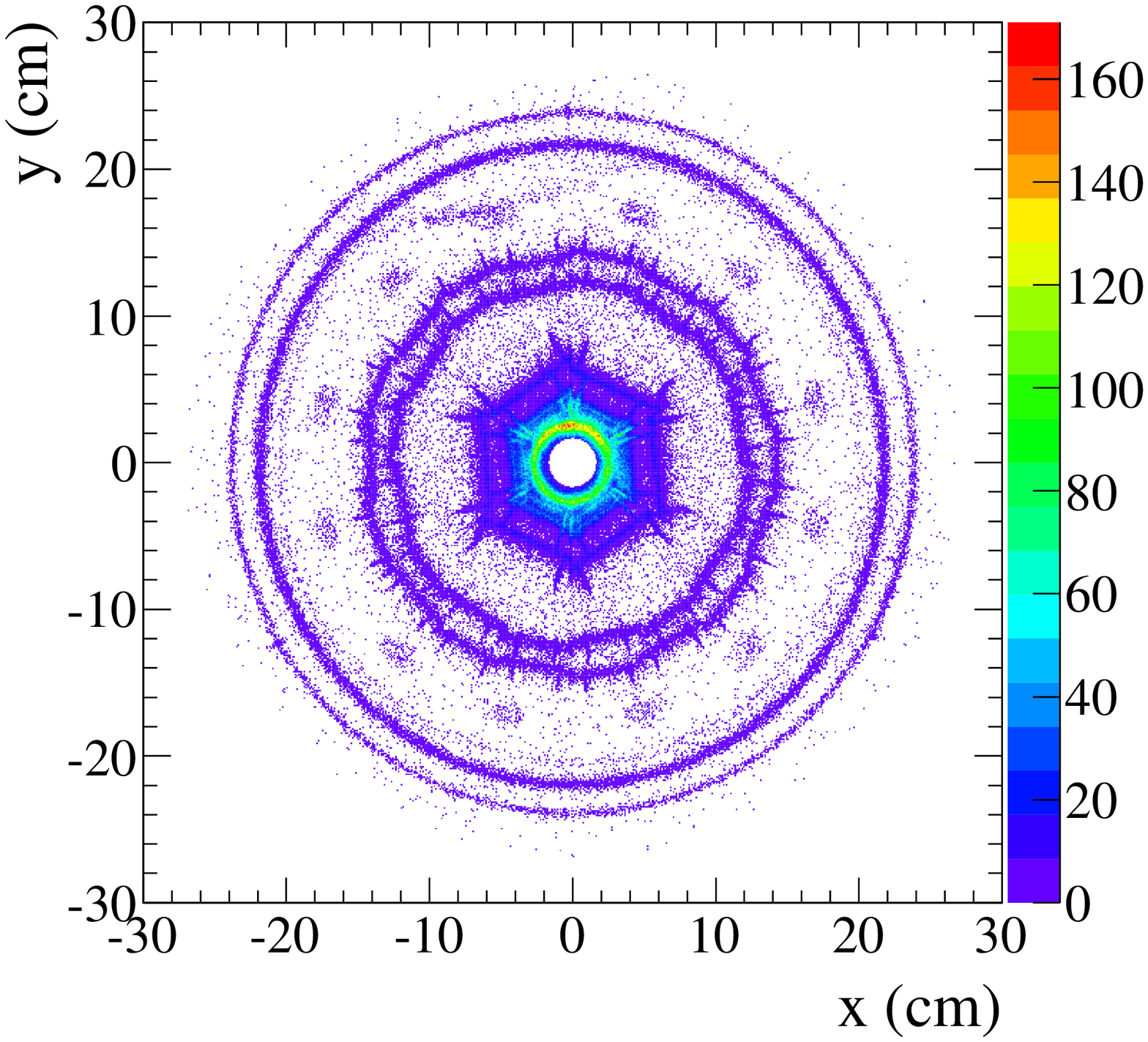}
\includegraphics[scale=0.42]{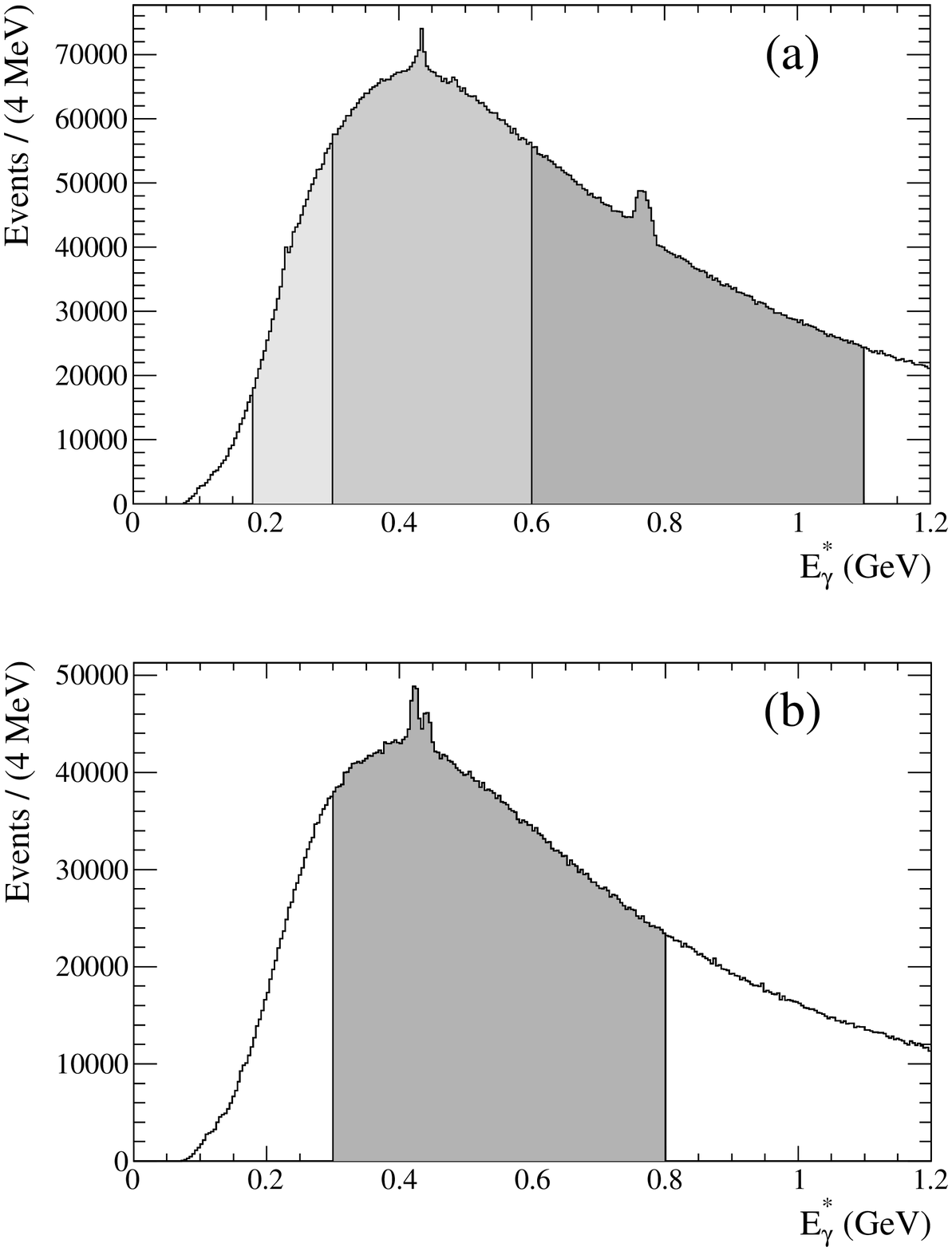}
  \begin{picture}(0.,0.)
    \put(-400,250){\bf \large BaBar Preliminary}
  \end{picture}
\caption{(left) End view of the the BaBar inner detector along the beam axis as seen by converted photons. Points indicate the number of converted photon vertices per cross-sectional area, as measured in a subset of the ``test sample'' data. 
From the center outwards, features of note include the beam pipe, the SVT (\emph{e.g.} hexagonal inner layers) and its support structure rods, the support tube, and the inner wall of the DCH. 
(right) The inclusive converted photon energy spectrum from (a) $\Upsilon(3S)$ and (b) $\Upsilon(2S)$ decays. The shaded areas indicate different regions of interest considered in detail in this analysis.}\label{fig_raw}
\end{center}
\end{figure}

The rate of $\gamma\rightarrow e^+e^-$ conversion in the detector material, and of the reconstruction of the $e^+e^-$ pairs is much 
lower than that for photons reconstructed using the BaBar electromagnetic calorimeter.  
However, the substantial improvement in energy resolution for photon conversions results in  
 better separation of photon energy lines.  

Photon conversions are reconstructed with a dedicated fitting algorithm that pairs oppositely charged particle tracks to form secondary vertices 
away from the interaction point. To remove internal conversions and Dalitz decays, and to improve signal purity, 
the conversion vertex radius is restricted to the beampipe, SVT, support tube, and inner wall of the DCH (see Fig.~1).  

The data are divided into four energy ranges (see Fig.~2) corresponding to bottomonium transitions, in $\Upsilon(3S)$ data,
\begin{itemize}
\item{$180\leq E^*(\gamma)\leq300$ MeV: $\chi_{bJ}(2P)\rightarrow\gamma\Upsilon(2S)$}
\item{$300\leq E^*(\gamma)\leq600$ Mev: $\Upsilon(3S)\rightarrow\gamma\chi_{bJ}(1P)$ and $\Upsilon(3S)\rightarrow\gamma\eta_{b}(2S)$}
\item{$600\leq E^*(\gamma)\leq1100$ MeV: $\chi_{bJ}(2P)\rightarrow\gamma\Upsilon(1S)$ and $\Upsilon(3S)\rightarrow\gamma\eta_{b}(1S)$}
\end{itemize}
and in $\Upsilon(2S)$ data,
\begin{itemize}
\item{$300\leq E^*(\gamma)\leq800$ MeV: $\chi_{bJ}(1P)\rightarrow\gamma\Upsilon(1S)$ and $\Upsilon(2S)\rightarrow\gamma\eta_{b}(1S)$.}
\end{itemize}

The photon energy ($E^*(\gamma)$) is analysed in the center-of-mass (CM) frame of the initial $\Upsilon(2S,3S)$ system.  The photon spectra from 
subsequent boosted decays (\emph{e.g.} $\chi_{bJ}(nP)\rightarrow\gamma\Upsilon(1S)$) are consequently 
affected by Doppler broadening due to the motion of the parent state in the CM frame.  
This broadening of the lineshapes of the various photon lines is incorporated into the fits used to extract the yields for the branching fractions obtained. 

The number of signal events for a given bottomonium transition is extracted from the data by 
performing a $\chi^{2}$ fit to the $E^*(\gamma)$ distribution in 1 MeV intervals.
The functional form and parameterization for each photon signal are determined from simulation.
In general, the lineshape is related to a Crystal Ball function \cite{ref:CB}, \emph{i.e.} a Gaussian function with a power-law tail.
This functional form is used to account for bremsstrahlung losses of the $e^+e^-$ pair.
The underlying smooth inclusive photon background is described by the product of a polynominal and an exponential function.
This functional form adequately describes the background in each separate energy range.

The decays $\Upsilon(3S)\rightarrow\gamma\chi_{b0,2}(1P)$ are observed, and precise measurements of the branching fractions
for $\chi_{b1,2}(1P,2P)\rightarrow\gamma\Upsilon(1S)$ and $\chi_{b1,2}(2P)\rightarrow\gamma\Upsilon(2S)$ decays obtained.
The background-subtracted fit results are shown in Fig.~2.   
The product of branching fractions $\mathcal{B}(\Upsilon(nS)\rightarrow\gamma\chi_{bJ})\times\mathcal{B}(\chi_{bJ}\rightarrow\Upsilon(1S))$ obtained
from the fit of  Fig.~2 are consistent with, and improve upon, the current values~\cite{ref:PDG}.

\begin{figure}  
\begin{center}
\includegraphics[scale=0.4]{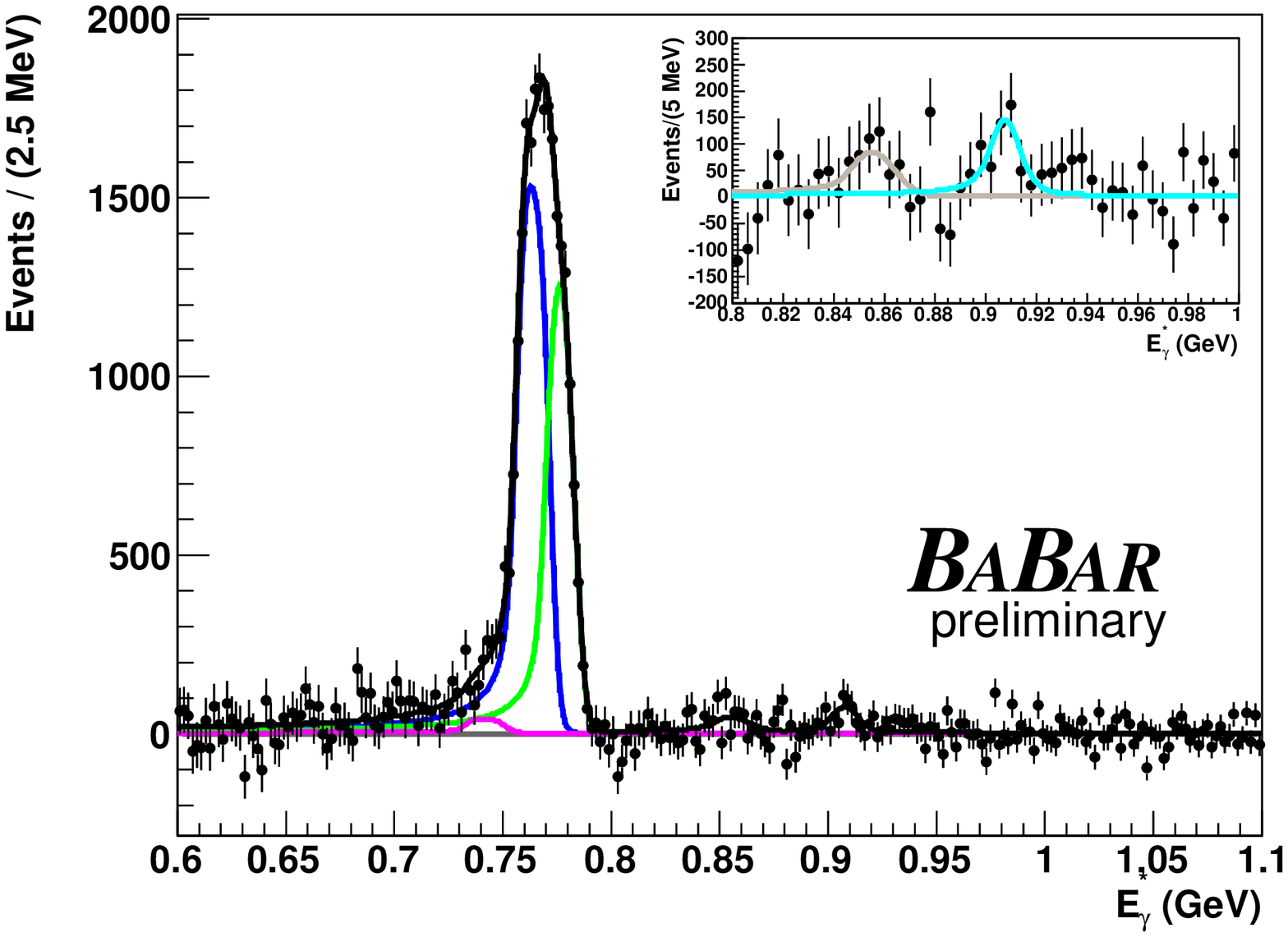}
\includegraphics[scale=0.4]{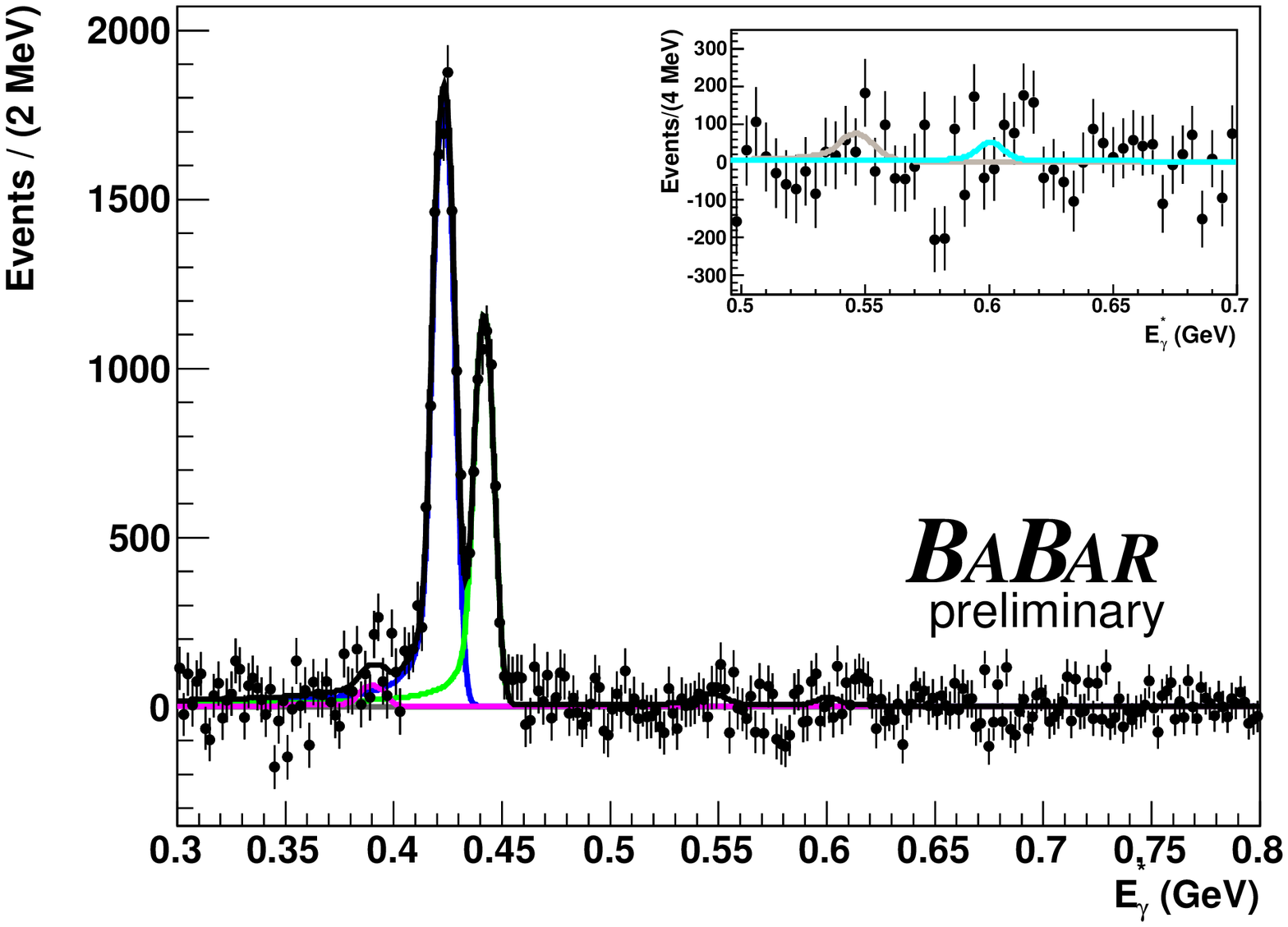}
\end{center}
\caption{Fit results for the inclusive converted photon spectrum after background-subtraction 
for (left) $\Upsilon(3S)$, and (right) $\Upsilon(2S)$ data. 
The purple, blue, green, gray, and cyan curves 
represent (left) $\chi_{b0,1,2}(2P)\rightarrow\gamma\Upsilon(1S)$ and (right) $\chi_{b0,1,2}(1P)\rightarrow\gamma\Upsilon(1S)$, initial state radiation (ISR), and $\Upsilon(nS)\rightarrow\gamma\eta_{b}(1S)$events, respectively.}
\end{figure}

\begin{table*}
\begin{center}
\begin{tabular}{cc}
\hline \hline
Transition & Branching Fraction Measurement $(\%)$\\
\hline
$\chi_{b0}(2P)\rightarrow\gamma\Upsilon(2S)$ & $-4.9\pm2.9^{+0.7}_{-0.8}\pm0.5 \:(<2.9)$ \\
$\chi_{b1}(2P)\rightarrow\gamma\Upsilon(2S)$ & $19.5\pm1.1^{+1.1}_{-1.0}\pm1.9$ \\
$\chi_{b2}(2P)\rightarrow\gamma\Upsilon(2S)$ & $8.6^{+0.9}_{-0.8}\pm0.5\pm1.1$ \\
\hline 
$\Upsilon(3S)\rightarrow\gamma\chi_{b2}(1P)$ & $10.6\pm0.3\pm0.6$ \\
$\Upsilon(3S)\rightarrow\gamma\chi_{b1}(1P)$ & $0.5\pm0.3^{+0.2}_{-0.1} \: (<1.1)$ \\
$\Upsilon(3S)\rightarrow\gamma\chi_{b0}(1P)$ & $2.7\pm0.4\pm0.2$\\
\hline
$\chi_{b0}(1P)\rightarrow\gamma\Upsilon(1S)$ & $2.3\pm1.5^{+1.0}_{-0.7}\pm0.2 \: (<4.6)$ \\
$\chi_{b1}(1P)\rightarrow\gamma\Upsilon(1S)$ & $36.2\pm0.8\pm1.7\pm2.1$  \\
$\chi_{b2}(1P)\rightarrow\gamma\Upsilon(1S)$ & $20.2\pm0.7^{+1.0}_{-1.4}\pm1.0$ \\
$\Upsilon(2S)\rightarrow\gamma\eta_b(1S)$ & $0.11\pm0.04^{+0.07}_{-0.05} \: (<0.22)$\\
\hline
$\chi_{b0}(2P)\rightarrow\gamma\Upsilon(1S)$ & $0.7\pm0.4^{+0.2}_{-0.1}\pm0.1 \:(<1.2)$ \\
$\chi_{b1}(2P)\rightarrow\gamma\Upsilon(1S)$ & $9.9\pm0.3\pm0.4\pm0.9$ \\
$\chi_{b2}(2P)\rightarrow\gamma\Upsilon(1S)$ & $7.1\pm0.2\pm0.3\pm0.9$ \\
$\Upsilon(3S)\rightarrow\gamma\eta_b(1S)$ & $0.059\pm0.016^{+0.014}_{-0.016}$\\
\hline \hline
\end{tabular}
\caption {Summary of Branching Fraction Measurements.  The first uncertainty is statistical and the second systematic. Upper limits are given at the $90\%$ confidence level.}
\end{center}
\end{table*}

The searches for $\eta_{b}(1S)$ and $\eta_{b}(2S)$ states using the converted photon energy spectrum are largely inconclusive.
Over a range of approximately $9974<m_{\eta_{b}(2S)}<10015$ MeV/$c^2$,
we find $\mathcal{B}(\Upsilon(3S)\rightarrow\gamma\eta_{b}(2S))<1.9\times 10^{-3}$ (at 90\% C.L.).
This value is consistent with, but does not improve upon, the upper limit obtained by CLEO~\cite{cleo_inclusive}.
Due to low efficiency and high background levels, no evidence for $\Upsilon(2S)\rightarrow\gamma\eta_b(1S)$ is found.
The most significant peaking structure seen in the $E^*(\gamma)$ energy region expected for the   
$\Upsilon(3S)\rightarrow\gamma\eta_b(1S)$ transition, if interpreted as an  $\eta_b$ signal,
trends toward recent potential
model~\cite{theory_etab_models} and lattice \cite{theory_etab_lattice} predictions.  However, the small ($<3\sigma$)
signicance of the result is insufficient to measure the $\eta_b$ mass from the present analysis.
Taking advantage of the improved resolution from a converted photon technique to make a
definitive measurement of the $\eta_b$ mass and width will require much more data from future experiments.  

The measured branching fractions obtained from the fits of Fig.~2 are listed in Table~1. 

\section{The BaBar search for the \boldmath$h_{b}(1P)$ in $\Upsilon(3S)$ data}

The $h_b(1P)$ state is expected to be produced in $\Upsilon(3S)$ decay via
$\pi^0$ or di-pion emission, and to undergo a subsequent $E1$
transition to the $\eta_b(1S)$, with branching fraction (BF)
${\mathcal B}(h_b(1P)\rightarrow \gamma \eta_b(1S))\sim (40-50)\%$~\cite{ref:Rosner2002}.
The isospin-violating
decay $\Upsilon(3S) \rightarrow \pi^0 h_b(1P)$ is expected to have a BF of about 0.1\%~\cite{ref:BFpredict,ref:Godfrey}, while
theoretical predictions for the transition $\Upsilon(3S) \rightarrow \pi^+\pi^- h_b(1P)$
range from $\sim 10^{-4}$~\cite{ref:BFpredict} up to $\sim 10^{-3}$~\cite{ref:standardmultipoleTuan,ref:Kuang,ref:kuangyan}.

We search for a signal 
in the inclusive recoil mass distribution against di-pion ($m_R$) or $\pi^0$ ($m_\mathrm{recoil}(\pi^0)$)
candidates.

\subsection{The search for the  \boldmath$h_{b}(1P)$ in the decay \boldmath$\Upsilon(3S)\rightarrow\pi^{+}\pi^{-} h_{b}$}

The $h_b$ signal is expected to appear as a peak in the $m_R$ distribution on top of a smooth non-peaking 
background from continuum events ($e^+e^- \to q\bar q$ with $q=u,d,s,c$) and
bottomonium decays.
Several other processes produce peaks in the recoil mass spectrum 
close to the signal region.
Hadronic transitions 
$\Upsilon(3S)\to\pi^+\pi^-\Upsilon(2S)$ 
(hereafter denoted $\Upsilon^{3\to 2}$) produce 
a peak centered at the $\Upsilon(2S)$ mass ($m[\Upsilon(2S)]=10.02326 \pm 0.00031$ GeV/$c^2$~\cite{ref:PDG}).
The cascade process 
$\Upsilon(3S)\to\Upsilon(2S)X$, 
$\Upsilon(2S)\to\pi^+\pi^-\Upsilon(1S)$ ($\Upsilon^{2\to 1}$)
results in a peak centered at $9.791$ GeV/$c^2$. 
The peak is offset from the $\Upsilon(1S)$ mass by approximately the 
$\Upsilon(3S)$ to $\Upsilon(2S)$ mass difference.
Doppler shift and broadening further affect the position and width of this peak.
When the $\Upsilon(2S)$ parent 
in $\Upsilon^{2\to 1}$ decays 
is produced through the $\Upsilon^{3\to 2}$ channel, 
a pion from the $\Upsilon(3S)$ decay can be combined with an 
oppositely-charged  
track from the $\Upsilon(2S)$ decay to produce a broad distribution 
centered around $9.9$ GeV/$c^2$.
The $\Upsilon(2S)$ is also produced through the 
initial-state radiation (ISR) process $e^+e^-\to \gamma_{ISR} \Upsilon(2S)$ 
($\Upsilon^{2\to 1}_{ISR}$). %
Of the nine possible $\Upsilon(3S)\to\chi_{bJ'}(2P)\gamma$, $\chi_{bJ'}(2P)\to\pi^+\pi^-\chi_{bJ}(1P)$ decay chains ($\chi_b^{J',J}$), 
only those for $J'=J=\{1,2\}$ have been reported~\cite{ref:PDG}; these should generate two slightly separated peaks near $9.993$ GeV/$c^2$, 
while the contributions with $J'\neq J$ or with $J=0$ are expected to be negligible. 

\begin{figure}[!hb]
\begin{center}
\includegraphics[scale=0.4]{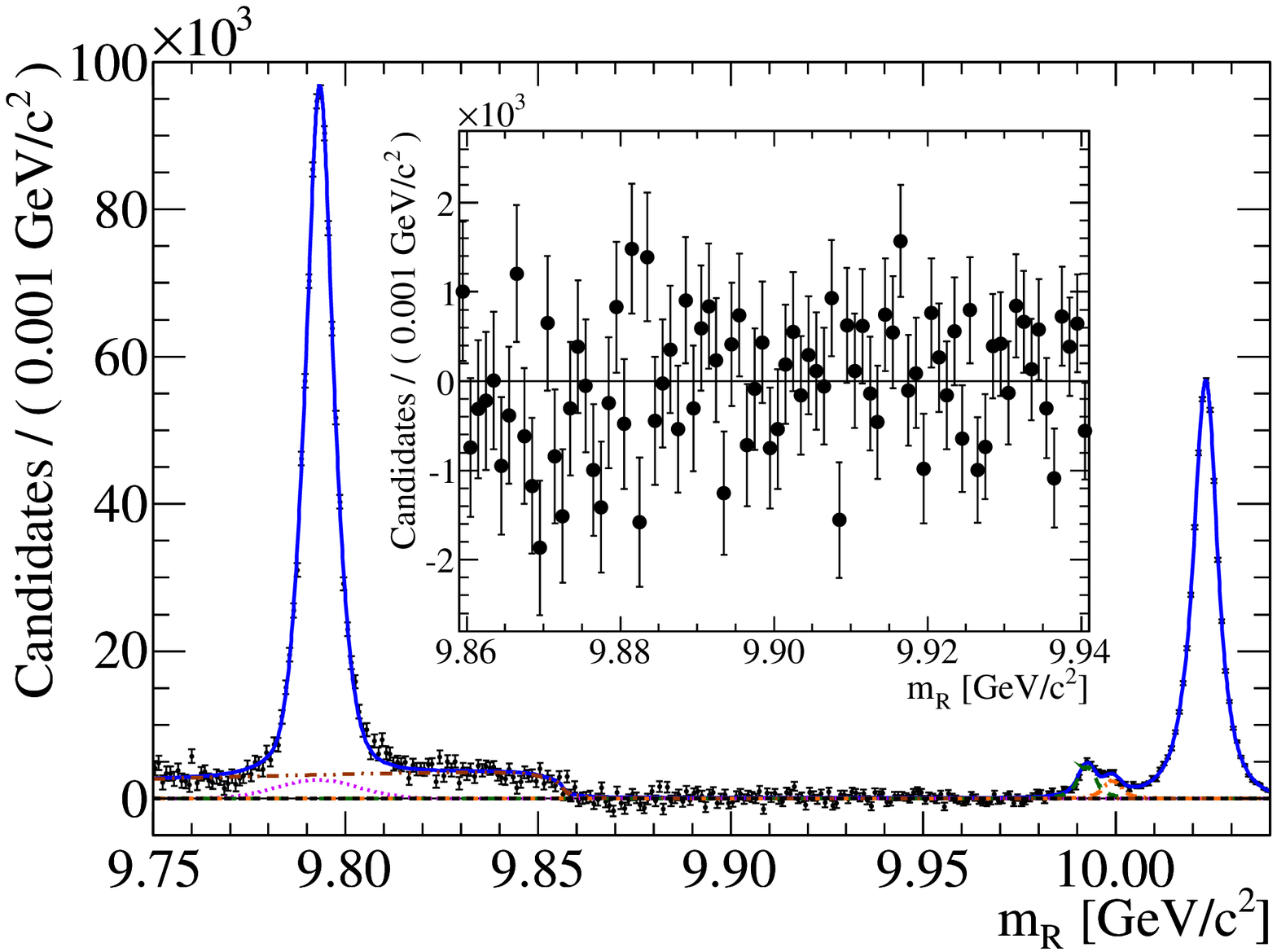}
\includegraphics[scale=0.4]{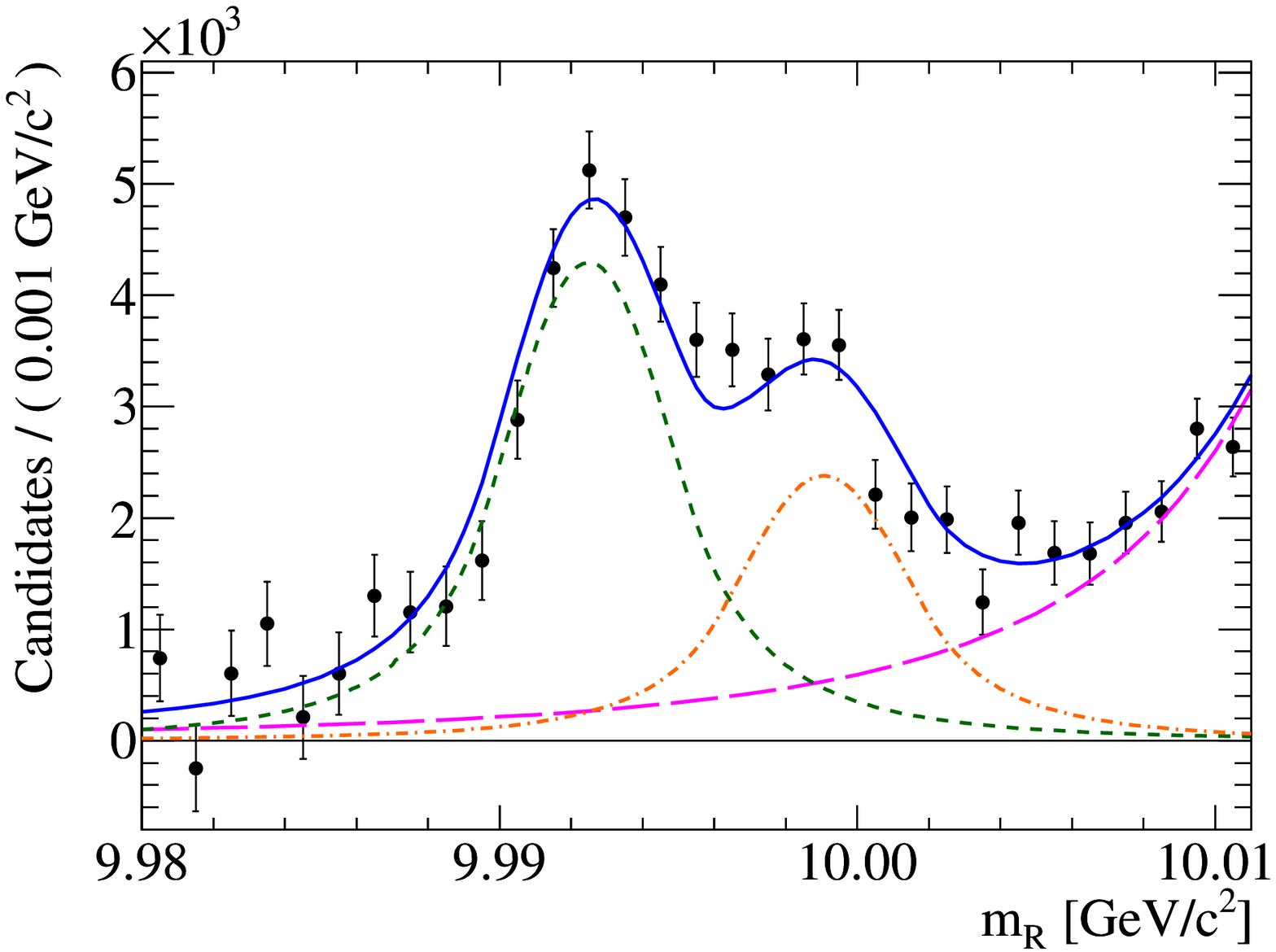}
\caption{(left) Results of the search for $\Upsilon(3S)\rightarrow\pi^{+}\pi^{-} h_{b}(1P)$. 
 The $m_R$ spectrum after subtraction of the continuum
  background component. 
The curves represent the
  $\Upsilon^{2\to 1}_{ISR}$ (dotted),
  $K_{S}^{0}$ (double-dot-dashed),
  $\chi_b^{1,1}$ (dashed),
  and $\chi_b^{2,2}$ (dot-dashed) components.
  Inset: expanded view in the $h_b$
region
  after subtraction of continuum and peaking backgrounds. (right) The $m_R$ spectrum in the $\chi_b^{J',J}$ region 
  after subtraction of continuum and $K_S$ background components: 
  points represent data, while the curves represent the fitted model (solid), the $\chi_b^{1,1}$ 
  (dashed), $\chi_b^{2,2}$ 
  (dot-dashed), and $\Upsilon^{3\to 2}$ 
  (long-dashed) components.
}
\end{center}
\end{figure}

No $h_b(1P)$ signal is seen in the $\Upsilon(3S)\rightarrow\pi^{+}\pi^{-} h_{b}$ production channel.  
Assuming the $h_b$ mass to be $9.900$ GeV/$c^2$, we set a 90\% C.L. upper limit 
${\cal B}_{\Upsilon}<1.2 \times 10^{-4}$.
We exclude, at 
90\% CL, values of ${\cal B}_{\Upsilon}$ above $1.8\times 10^{-4}$ 
for a wide range of assumed $h_b$ mass values. 
These results disfavor the calculations 
of Refs.~\cite{ref:standardmultipoleTuan,ref:Kuang,ref:kuangyan}. 
Similarly, a recent measurement of the 
$\Upsilon(1^3D_J)\to\Upsilon(1S)\pi^+\pi^-$ branching fraction~\cite{ref:bab1d}
disfavors the calculations of Ref.~\cite{ref:Kuang,ref:kuangyan}.
The predictions of Ref.~\cite{ref:Voloshin} are 
at least one order of magnitude smaller and are not contradicted by our result.

Figure~3 shows  the distribution of $m_R$ 
after subtraction of the non-peaking background.   
An expanded view of the $\chi_b^{J',J}$ region is shown on the right.

A branching fraction measurement 
${\cal B}[\Upsilon(3S)\to\pi^+\pi^-\Upsilon(2S)]=(3.00\pm 0.02{\rm (stat.)}\pm 0.14{\rm (syst.)})\% $
 improves over  
 the current world average $(2.45\pm 0.23)\%$~\cite{ref:PDG}.
Product branching fractions 
${\cal B}[\Upsilon(3S)\to X\chi_{b1}(2P)]\times {\cal B}[\chi_{b1}(2P)\to\pi^+\pi^-\chi_{b1}]= (1.16\pm 0.07\pm 0.12)\times 10^{-3}, $
${\cal B}[\Upsilon(3S)\to X\chi_{b2}(2P)]\times {\cal B}[\chi_{b2}(2P)\to\pi^+\pi^-\chi_{b2}]
= (0.64\pm 0.05\pm 0.08)\times 10^{-3}, $ and
${\cal B}[\Upsilon(3S)\to X\Upsilon(2S)]\times {\cal B}[\Upsilon(2S)\to\pi^+\pi^-\Upsilon]= (1.78\pm 0.02\pm 0.11)\% $
 are also obtained from this analysis. 
In addition, we obtain a measurement of $331.50 \pm 0.02 ({\rm stat.}) \pm 0.13({\rm syst.})$ MeV/$c^2$ for the $\Upsilon(3S)$-$\Upsilon(2S)$ 
mass difference.

\subsection{Evidence for the  \boldmath$h_{b}(1P)$ in the decay \boldmath$\Upsilon(3S)\rightarrow\pi^{0} h_{b}$}

Evidence for the $h_b(1P)$ state is obtained in the decay $\Upsilon(3S)\rightarrow \pi^{0} h_b(1P)$, by requiring a photon 
with an energy consistent with that for the $h_{b}(1P)\rightarrow\gamma\eta_{b}(1S)$ transition.  
The number of $\pi^{0}$ events in $m_\mathrm{recoil}(\pi^{0})$ is determined from a fit to the $m_{\gamma\gamma}$ distribution in each 
$m_\mathrm{recoil}(\pi^{0})$ interval.

A $\pi^0$ candidate is reconstructed as a photon pair with invariant mass $m(\gamma \gamma)$ in the range 55--200 MeV/$c^2$ 
(see Fig.~4).  
In the calculation of $m_\mathrm{recoil}(\pi^0)$, the $\gamma$-pair invariant mass is 
constrained to the nominal $\pi^0$ value~\cite{ref:PDG} in order to improve the momentum resolution of the $\pi^0$.
We employ a simple set of criteria based on kinematic and event shape variables to suppress backgrounds. 
Photons from $\pi^0$ decays are a primary source of background in the region of the signal photon line from $h_b\rightarrow \gamma \eta_b$ 
transitions.  Similarly, many misreconstructed $\pi^0$ candidates result from the pairing of photons from different 
$\pi^0$'s.  We impose the $\pi^0$ veto\footnote{A signal photon candidate is rejected if, when combined with another photon
in the event ($\gamma_2$), the resulting $\gamma \gamma_2$ invariant mass is within 15 MeV/$c^2$ of the nominal $\pi^0$ mass;
this is called a $\pi^0$ veto.} condition on all photon candidates to improve the purity of the $\pi^0$ and photon candidate sample. 

We obtain the $m_\mathrm{recoil}(\pi^0)$ distribution in 90 intervals of 3 MeV/$c^2$ from 9.73 to 10 GeV/$c^2$. 
For each $m_\mathrm{recoil}(\pi^0)$ interval, the $m(\gamma \gamma)$ spectrum consists of a $\pi^0$ signal above combinatorial 
background (see Fig.~4). 
We construct the $m_\mathrm{recoil}(\pi^0)$ spectrum by extracting the $\pi^0$ signal yield in each interval of $m_\mathrm{recoil}(\pi^0)$ 
from a fit to the $m(\gamma\gamma)$ distribution in that interval.     
The $m_\mathrm{recoil}(\pi^0)$ distribution is thus obtained as the fitted $\pi^0$ yield and its uncertainty for each interval of $m_\mathrm{recoil}(\pi^0)$.

\begin{figure}[ht]
\begin{center}
 \includegraphics[width=.55\textwidth]{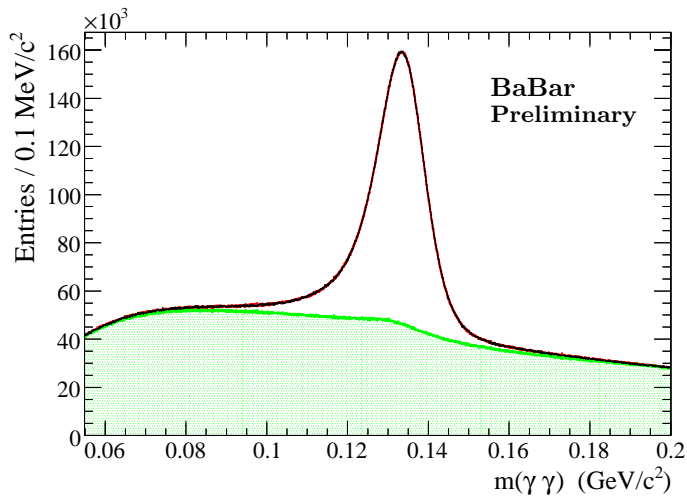}
  \begin{picture}(0.,0.)
    \put(-85,160){\bf \normalsize BaBar}
    \put(-85,150){\bf Preliminary}
  \end{picture}
  \caption{The result of the fit to the $m(\gamma \gamma)$ distribution in data (data points) for the full range of 
$m_\mathrm{recoil}(\pi^0$).  The solid 
histogram shows the fit result, and is essentially indistinguishable from the data; the shaded histogram
corresponds to the background distribution.  }
\end{center}
\end{figure}

We use the MC background and MC $\pi^0$-signal distributions 
directly in fitting the $m(\gamma\gamma)$ distributions in data. 
For each $m_\mathrm{recoil}(\pi^0)$ interval in MC, we obtain histograms in 0.1 MeV/$c^2$ intervals of $m(\gamma \gamma)$ corresponding to 
the $\pi^0$-signal and background distributions.  The $\pi^0$-signal distribution is obtained by 
requiring matching of the reconstructed to the generated $\pi^0$'s on a candidate-by-candidate basis 
(termed ``truth-matching'' in the following discussion).  
The histogram representing background is obtained by subtraction of the $\pi^0$ signal from the total distribution. 

For both signal and 
background the qualitative changes in shape over the full range of $m_\mathrm{recoil}(\pi^0)$ are quite well reproduced 
by the MC.  
However, the $\pi^0$ signal distribution in data is slightly broader than in MC, and is peaked at a slightly higher mass value.    
The $m(\gamma \gamma)$  background shape also differs between data and MC.  
To address these differences, the MC $\pi^0$ signal is displaced in mass and smeared by a double Gaussian 
function with different mean and width values;  
the MC background distribution is weighted according to a polynomial in
$m(\gamma \gamma)$.  
The signal-shape and background-weighting parameter values are obtained from a fit to the $m(\gamma \gamma)$ distribution in data
for the full range of $m_\mathrm{recoil}(\pi^0)$.  
At each step
in the fitting procedure, the $\pi^0$ signal and background distributions
are normalized to unit area, and a $\chi^2$ between a linear combination of these MC histograms and the $m(\gamma \gamma)$ distribution in data is computed. 
The fit function provides an excellent description of the data ($\chi^{2}/NDF$=1446/1433; $NDF$=Number of Degrees of Freedom) 
and the fit result is essentially indistinguishable from the data histogram.  
The background distribution exhibits a small peak at the
$\pi^0$ mass, due to interactions in the detector material of
the type $n \pi^+ \rightarrow p \pi^0$ or $p \pi^- \rightarrow n \pi^0$ that cannot be truth-matched. The normalization of this
background to the non-peaking background is obtained from the MC simulation, which incorporates the
results of detailed studies of interactions in the detector material
performed using data~\cite{ref:geant}. This peak is displaced and smeared as for the
primary $\pi^0$ signal.

 The fits to the individual
$m(\gamma \gamma)$ distributions are performed with the smearing and weighting parameters fixed to the values
obtained from the fit shown in Fig.~4.  In this process, the MC signal and
background distributions for each $m_\mathrm{recoil}(\pi^0)$ interval are shifted,
smeared, and weighted using the fixed parameter values, and then
normalized to unit area.  Thus, only the signal and background yields are  free parameters
in each fit. 
The $\chi^2$ fit to the data then gives the value and
 the uncertainty of the number of $\pi^0$ events in each $m_\mathrm{recoil}$
 interval.  

\begin{figure}
\begin{center}
\includegraphics[width=.45\textwidth]{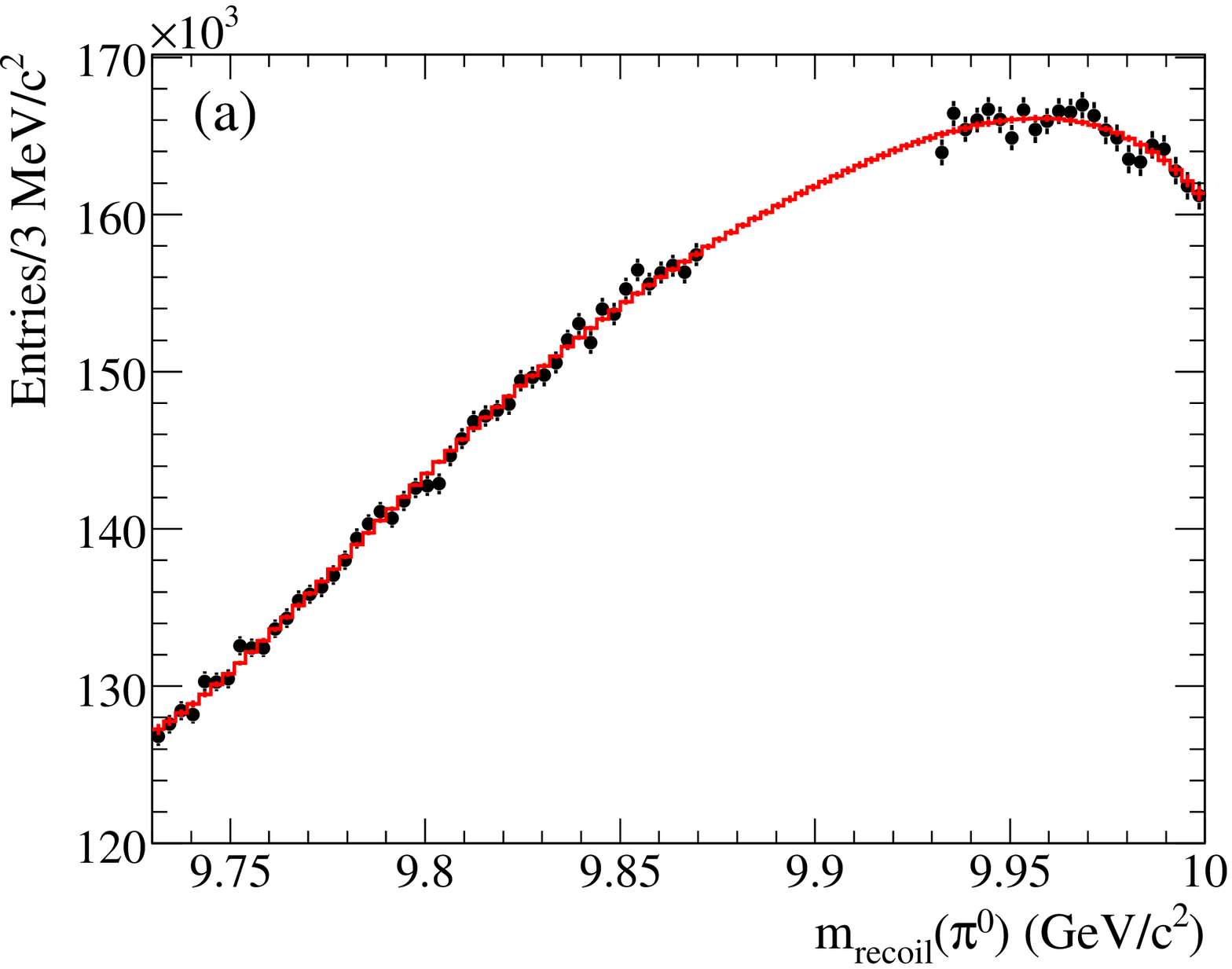}
\includegraphics[width=.45\textwidth]{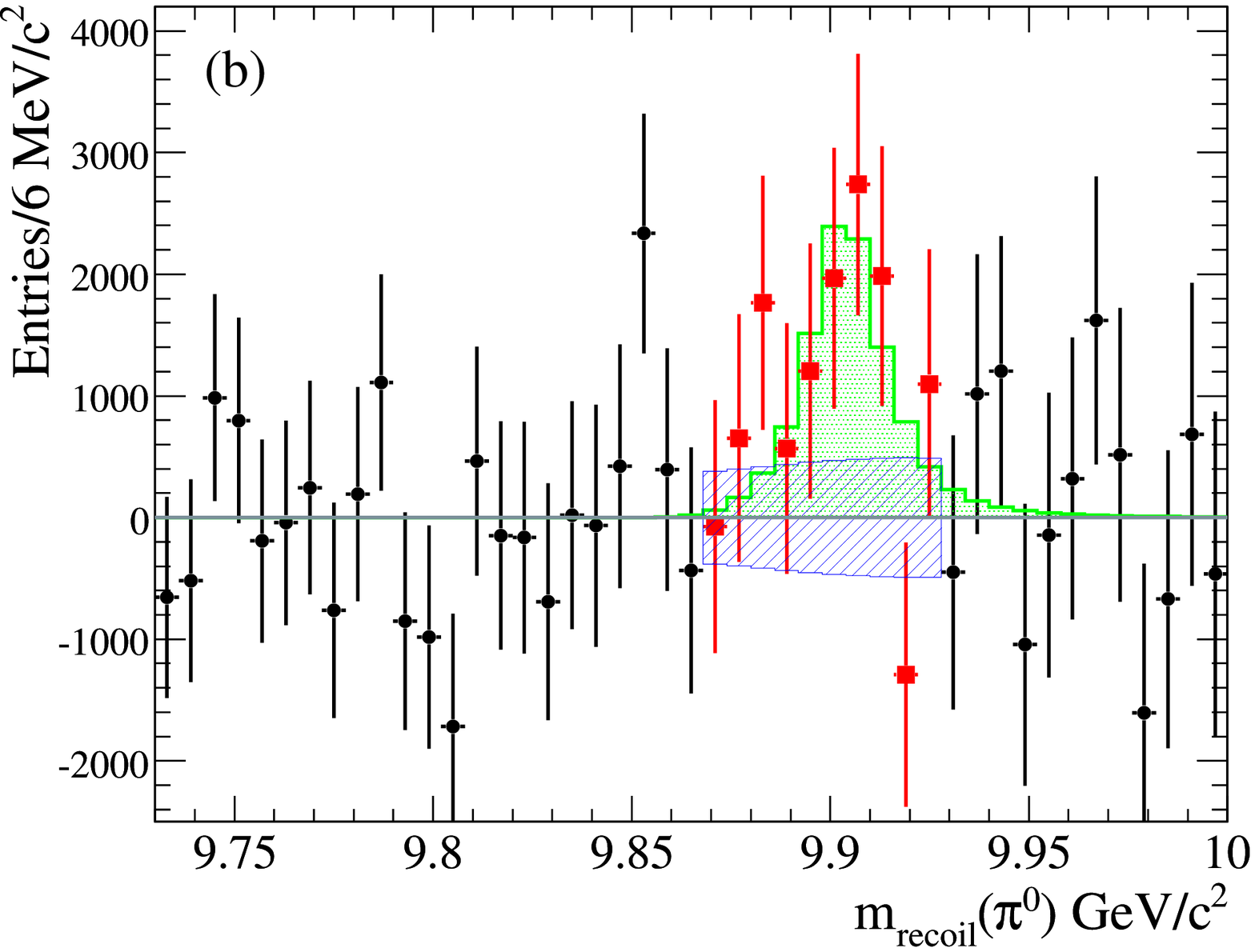}
  \begin{picture}(0.,0.)
    \put(-390,150){\bf \normalsize BaBar}
    \put(-390,140){\bf Preliminary}
    \put(-170,150){\bf \normalsize BaBar}
    \put(-170,140){\bf Preliminary}
  \end{picture}

  \caption{ (a) The $m_\mathrm{recoil}(\pi^0)$ distribution in the region $9.73<m_\mathrm{recoil}(\pi^0)<10$ GeV/$c^2$ for data (points);
the solid histogram represents the fit function described in the text.
The data in the $h_b$ signal region have been excluded from the fit and the plot.
(b) The $m_\mathrm{recoil}(\pi^0)$ spectrum after subtracting background; in the $h_b$ signal region the data points are shown as squares,
and the area with diagonal shading represents the uncertainties from the background fit;
the shaded histogram represents the signal function resulting from the fit to the
data in the signal region.}
\end{center}
\end{figure}

To search for an $h_b$ signal, we perform a binned $\chi^{2}$ fit to the $m_\mathrm{recoil}(\pi^0$) distribution obtained in data.
The $h_b$ signal function is represented by the sum of two
Crystal Ball functions~\cite{ref:CB} with parameter values, other than the $h_b$ mass, $m$, and the normalization,
 determined from simulated signal $\Upsilon(3S)\rightarrow \pi^0 h_b$ events.  The background function is a fifth order polynomial, 
where parameter values are obtained from a fit which excludes the signal region (see Fig.~5(a)).

We then perform a fit over the signal region to search for an $h_b$ signal of the expected
shape, taking account of the correlated uncertainties related to the polynomial interpolation procedure, but with the fitted values of the 
polynomial coefficient fixed.

In Fig.~5(b) we plot the difference between the distribution of $m_\mathrm{recoil}(\pi^0)$ and the fitted histogram of
Fig.~5(a) over the entire region from 9.73 GeV/$c^2$ to 10.00 GeV/$c^2$; we have combined pairs of 3 MeV/$c^2$ intervals from
Fig.~5(a) for clarity.

The width of the observed signal is
consistent with experimental resolution, and its significance is 3.3 $\sigma$, including systematic uncertainties.
The measured mass value, $m=9902\pm 4$(stat.)$\pm 2$(syst.) MeV/$c^2$,  
 is consistent with the expectation for the $h_b(1P)$ bottomonium state,
the axial vector partner of the $\chi_{bJ}(1P)$ triplet of states.
We obtain the value $(4.3 \pm 1.1 \rm{(stat.)} \pm 0.9 \rm{(syst.)} )\times 10^{-4}$ for the product branching fraction
${\mathcal B}(\Upsilon(3S)\rightarrow \pi^0 h_b)\times {\mathcal B}(h_b\rightarrow \gamma \eta_b)$.

This measurement combined with the expected rate for 
the $h_{b}(1P)\rightarrow\gamma\eta_{b}(1S)$ transition~\cite{ref:Rosner2002} would 
indicate that the $\Upsilon(3S)\rightarrow\pi^{+}\pi^{-} h_{b}(1P)$ is suppressed by a factor greater than 3 w.r.t. 
$\Upsilon(3S)\rightarrow\pi^{0} h_{b}(1P)$.

\section{Conclusions}

A study of radiative transitions between bottomonium states using photons that have been 
converted to $e^+e^-$ pairs in the detector material led to the observation of 
 $\Upsilon(3S)\rightarrow\gamma\chi_{b0,2}(1P)$ decays, 
and to precise measurements of the branching fractions for $\chi_{b1,2}(1P,2P)\rightarrow\gamma\Upsilon(1S)$ and $\chi_{b1,2}(2P)\rightarrow\gamma\Upsilon(2S)$ decays.

Studies of inclusive di-pion and $\pi^0$ transitions was carried out using 
$\Upsilon(3S)$ events.
While we find no evidence for the bottomonium spin-singlet state $h_b(1P)$ in the 
invariant mass distribution recoiling against the $\pi^+\pi^-$ system, we obtain evidence at the 3.3$\sigma$ level for the $h_b(1P)$ 
in $\Upsilon(3S)\to \pi^0 h_b(1P)$ decay.  Shortly after the BaBar $h_b$ search results were obtained, 
preliminary results of a search for the $h_b$ in the reaction 
$e^+e^-\rightarrow h_b(nP) \pi^+\pi^-$ ($n=1,2$) in data collected near the $\Upsilon(5S)$ resonance 
were announced by the Belle Collaboration~\cite{ref:Belle}. The $h_b(1P)$ mass measured therein agrees very well with 
the value reported in this paper. It is interesting to note that the $e^+e^- \rightarrow h_b(nP)\pi^+\pi^-$ cross sections measured in the Belle analysis 
relative to the $e^+e^-\rightarrow \Upsilon(2S)\pi^+\pi^-$  cross section are large, 
indicating an anomalous production rate for the $h_b(1P)$ and $h_b(2P)$, 
and suggesting that these states are maybe produced via an exotic process that violates the
suppression of heavy quark spin-flip.

\end{document}